\documentclass[a4paper,11pt]{article}
\pdfoutput=1
\usepackage{jheppub1}
\usepackage[T1]{fontenc}
\usepackage{amssymb}
\usepackage{graphicx}
\usepackage{amsmath}
\usepackage{hyperref}
\usepackage{tensor}
\usepackage{epstopdf}
\usepackage{extarrows}
\usepackage{float}
\usepackage{tikz}
\usepackage{subcaption}
\usepackage{amsmath}
\newcommand{\Fi}{\text{Fi}}
\usepackage{changes}
\usepackage{verbatim}
\definechangesauthor[name={Per cusse}, color=orange]{per}

\begin{document}
	\title{Holographic quantum distances and replica trick}
	\author{Zi-Qing Xiao and Run-Qiu Yang}
	\emailAdd{kingshaw@tju.edu.cn}
\emailAdd{aqiu@tju.edu.cn}
	\affiliation{Center for Joint Quantum Studies and Department of Physics, School of Science, Tianjin University, Yaguan Road 135, Jinnan District, 300350 Tianjin, P.~R.~China}
	\abstract{This paper gives concrete examples to exhibit how to use the replica trick to calculate the quantum (quasi-)distances holographically. First, we consider the fidelity and relative entropy between thermal states that are dual to the Schwarzschild-AdS black holes. Then we generalize our method into the RN-AdS black holes by adding a U(1) gauge field. We also investigate the fidelity between states excited by scalar operator in probe limit. In this case, it is surprising that the fidelity in standard quantization will suffer from new UV divergence though the usual holographic renormalization has been applied. We call for deep understanding for such divergence in the future. We also discover a holographic method to check whether the density matrices of two holographic states are commutative. }
\maketitle
\noindent
\section{Introduction}
In recent years, quantum
information theoretic considerations play an important role in the study of AdS/CFT or more general gauge/gravity duality~\cite{Maldacena:1997re, Gubser:1998bc,Witten:1998qj}. For instance,	Ryu-Takayanagi (RT) formula~\cite{Ryu:2006bv, Hubeny:2007xt} quantifies the entanglement entropy of the boundary conformal field theories (CFTs), which is given by the area of a codimension-2 minimal surface in the dual bulk spacetime. Under some reasonable assumptions, RT formula can be obtained from the more general R\'{e}nyi entropy~\cite{Fursaev:2006ih, Headrick:2010zt, Lewkowycz:2013nqa, Nakaguchi:2016zqi}
$$
	S_n\equiv\frac{1}{1-n}\ln\frac{\text{tr}\rho^n}{(\text{tr}\rho)^n}
$$
by applying the \emph{replica trick}. Here the index $n$ labels the number of the replicated boundary CFTs and $\rho$ is the density matrix of the entangling region. The $n$ is originally assumed to be a positive integer value. In~\cite{Lewkowycz:2013nqa}, Lewkowycz and Maldacena argued that, for integer index $n>1$, the CFTs on a branched cover\footnote{See Ref.~\cite{Callan:1994py,Holzhey:1994we} for how to construct $\mathcal{M}_n$ by taking $n$ copies of $\mathcal{M}_1$. The key points here are making suitable identifications for quantum fields between these copies and obeying right boundary conditions. The latter, in other words, requires us to regulate the path integral of $\mathcal{M}_n$ by imposing the right field configurations at the boundaries.} $\mathcal{M}_n$ over the entangling region still have corresponding interior gravity duals $\mathcal{B}_n$ like the original Euclidean spacetime $\mathcal{B}_1$ does. In the gravity side, there exists a solution $B_n$ that satisfies the Einstein equation along with boundary condition $\partial \mathcal{B}_n=\mathcal{M}_n$. Based on the holographic duality, one should expect that the partition functions of two sides' theories are equated each other
\begin{equation}\label{dual}
	Z[\mathcal{M}_n]=Z[\mathcal{B}_n]\,,
\end{equation}
where $Z[\mathcal{M}_n]=\text{tr}\rho^n$. Note that we are working in the large $N$ limit such that the bulk geometry can be taken as a solution of Einstein gravity. Under the saddle point approximation, the partition function of bulk geometry can be written as the exponential function of corresponding Einstein-Hilbert action
\begin{equation}\label{sad}
	Z[\mathcal{M}_n]=Z[\mathcal{B}_n]\approx\exp(-I_{\text{bulk}}[\mathcal{B}_n])\,.
\end{equation}
Following R\'{e}nyi entropy's definition, the direct calculation
$$
	S_n=\frac{1}{1-n}\ln\frac{\text{tr}\rho^n}{(\text{tr}\rho)^n}=\frac{1}{n-1}\left(I_{\text{bulk}}[\mathcal{B}_n]-nI_{\text{bulk}}[\mathcal{B}_1]\right)
$$
indicates that R\'{e}nyi entropy is proportional to the gravitational action of corresponding bulk geometry. If it is possible to analytically continue the R\'{e}nyi entropy $S_n$ to non-integer\footnote{The Carlson's theorem shows that such analytical continuation is unique if it exists.} $n$, the von Neumann entropy that is usually referred as the entanglement entropy in quantum information theory can be recovered from $S_n$ at the $n\rightarrow1$ limit.
The entanglement entropy concerns about the entanglement property of only one quantum state but not involves the relationship between two states. There are other quantities in quantum information~\cite{nielsen_chuang_2010}, which measures how close two quantum states are. In other words, these quantities label some kind of ``distance'' between quantum states. For example, given quantum states $\rho$ and $\sigma$, the \emph{trace distance} is defined as
\begin{equation}
	D[\rho, \sigma]\equiv\frac{1}{2}\text{tr}\left|\rho-\sigma\right|\,,
\end{equation}
where $\text{tr}|A|\equiv\sum_{i}\lambda_i$ and $\lambda_i$ is the eigenvalue of $\sqrt{A^{\dagger}A}$. Introducing a positive number $n$, $D[\rho, \sigma]$ can be generalized into $D_n[\rho, \sigma]$
\begin{equation}
	D_n[\rho, \sigma]\equiv\frac{1}{2^{1/n}}\sqrt[n]{\text{tr}\left|\rho-\sigma\right|^n}\,.
\end{equation}
For $n=1$, the above formula is just the trace distance. There is other special choice that is widely applied. By choosing $n=2$, \emph{Hilbert-Schmidt distance} can be obtained, which is most convenient $D_n$ for calculations. The second families of quantum distance are based on the \emph{fidelity}
\begin{equation}
	\Fi[\rho, \sigma]\equiv\text{tr}\sqrt{\rho^\frac{1}{2}\sigma\rho^\frac{1}{2}}\,.
\end{equation}
The fidelity is not a distance but it can be used to define other measures of distance: the\emph{Fubini-Study distance} $D_F[\rho, \sigma]$ and the \emph{Bures distance} $D_B[\rho, \sigma]$
\begin{equation}
	D_F[\rho, \sigma]\equiv\arccos \Fi[\rho, \sigma],~~~D_B[\rho, \sigma]\equiv\sqrt{1-\Fi[\rho, \sigma]^2}\,.
\end{equation}
Due to the well-defined calculations in finite dimensional Hilbert spaces and the preservation under unitary transformations, trace distance and fidelity are widely used within the quantum information community. However, their analytical calculations are extremely difficult in quantum field theory. Ref.~\cite{Lashkari:2014yva} makes the first step towards this issue. By virtue of the certain correlation functions for the vacuum or the thermal state, author of Ref.~\cite{Lashkari:2014yva} develop a replica trick to calculate the fidelity for $(1+1)$ dimensional CFTs. Using similar method, Refs.~\cite{Zhang:2019wqo,Zhang:2019itb,Zhang:2019kwu} also developed replica method to compute the trace distance for a class of special states for single short interval in 1+1 dimensional CFTs. Recently,  Ref. ~\cite{Kudler-Flam:2022zgm} proposes how to use replica trick to compute R\'{e}nyi mutual information in 1+1 CFTs. Though in higher dimensions the replica trick is still effective, the correlation functions of interacting theories are highly nontrival to compute. Similar to holographic descriptions of entanglement, one would wish there exist a holographic duality to calculate trace distance and fidelity. Actually Ref.~\cite{Miyaji:2015woj} argues that the fidelity between a state and its infinitesimal perturbation state is approximately given by a volume of maximal time slice in AdS spacetime. However, when the difference between two states is not infinitesimal, the gravity dual of fidelity given by Ref.~\cite{Miyaji:2015woj} is not reliable. In addition, the holographic proof on the proposal of Ref.~\cite{Miyaji:2015woj} is absent. See~\cite{Colin-Ellerin:2020mva,Dong:2021clv,Colin-Ellerin:2021jev,Dong:2021oad,Penington:2022dhr} for recent progress related to replicas in quantum information under the holographic setup.
Although it seems quite difficult to  solve this problem completely, this paper uses some concrete examples to exhibit how to calculate quantum distances in holography. As a basic perspective in holographic picture, Maldacena argues that the thermal states in the boundary CFT correspond to Schwarzschild-AdS black holes. Firstly, similar to the deduction of R\'{e}nyi
entropy by employing the replica trick, we analytically calculate the fidelity and relative entropy between thermal states in holography. Next, we want to check our method in the presence of bulk matters. Here we consider two concrete examples. In the first one we consider that there is a conserved charge in the boundary theory so the bulk theory is the Einstein-Maxwell theory. In the second example,  an Einstein-scalar theory is constructed here to gain the insight into the fidelity between states excited by scalar operator. In probe limit, we derive the analytic expression of fidelity in this case. Surprisingly, we find that the fidelity of probe limit is well defined only in the alternative quantization. The fidelity in standard quantization will suffer from new UV divergency though the usual holographic renormalization has been applied. To the best of our knowledge, such kind of divergency has not been discussed well in the study of holography, and so we call for deep understanding in the future.
In a physical system described by quantum mechanics, quantum states are generally the superposition of several eigenstates under a observable Hermitian operator. It is more troublesome but feasible to check the commutativity of two quantum mechanics states. As we claimed before, quantum states in infinite dimensional Hilbert spaces are more tricky. Without more assumptions, it is difficult to directly write the density matrix for a general quantum state in quantum field theory. So it is nontrivial question to determine the commutativity of two quantum states in quantum field theory. At the end of this paper, we propose a method to answer this question partially if two quantum states are holographic. We will see whether two holographic states are commutative or not is related the contribution of corresponding bulk theory's action.

The organization of this paper is as follows. In section \ref{s2}, we consider the fidelity and relative entropy between thermal states with different temperatures. In section \ref{s3}, under the same
procedure, we generalize the above method into the grand canonical ensemble, where we consider the fidelity and relative entropy of the states which has different chemical potentials. In section \ref{s4}, we construct an Einstein-scalar theory in asymptotically AdS spacetime which is dual to states excited by scalar operator, and derive the semi-analytic formula of fidelity in the probe limit. Particularly, in the standard quantization, we show that the fidelity will suffer from a kind of new UV divergence even though we have added usual counterterms into the action. We will also show that such kind of new divergence will not appear in alternative quantization. In the section \ref{sec5}, we will discuss how to check the commutativity of two density matrices holographically.

\section{Holographic quantum distance between thermal states}\label{s2}
In this section, we show how to holographically get the quantum distances between thermal states. In addition, we will explain the main idea about how to use replica trick to holographically compute the expression such as tr$(\rho_1\rho_2\cdots\rho_n)$ for a series of given holographic states $\{\rho_1,\rho_2,\cdots,\rho_n\}$.

Our starting point is that thermal states in $\text{CFT}_d$ are dual to $\text{Schwarzschild-AdS}_{d+1}$ black hole~\cite{Maldacena:2001kr}. Let us start from the properties of the thermal state. Thermal state can be used to describe the quantum system which is in contact with a heat bath at temperature $T=1/\beta$. In the canonical ensemble at temperature $T$, the density matrix of thermal state reads
\begin{equation}
	\rho=\frac{e^{-\beta H}}{Z(\beta)}\,.
\end{equation}
Here $H$ is the Hamiltonian of this system and $Z(\beta)=\text{tr}e^{-\beta H}$ is the partition function. To illustrate how use replica trick to compute the quantum distance here, we will consider two thermal systems, of which the density matrices are
\begin{equation}\label{rho12}
	\rho_1=\frac{e^{-\beta_1 H}}{Z(\beta_1)}\,,~\rho_2=\frac{e^{-\beta_2 H}}{Z(\beta_2)}\,.
\end{equation}
In following we will first introduce how to compute the trace of product $\rho_1\rho_2$ holographically,
\begin{equation}
	\rho_1\rho_2=\frac{e^{-(\beta_1+\beta_2)H}}{Z(\beta_1)Z(\beta_2)}\,.
\end{equation}
This is product will play the essential role in our computations of quantum distances later on.

\begin{figure}[H]
	\centering
	\begin{tikzpicture}[thick, scale=1]
		\filldraw[draw=white!80,fill=gray!20] (1,1) circle (1);
		\draw[red] (2,1) arc (0:360:1);
		\draw[black] (1,1) node{$\rho_1$};
		\filldraw[draw=white!80,fill=gray!20] (3.5,1) circle (0.5);
		\draw[densely dashed,blue] (4,1) arc (0:360:0.5);
		\draw[black] (3.5,1) node{$\rho_2$};
		\draw[black] (2.5,1) node{$\times$};
		\draw[black] (4.5,1) node{$=$};
		\filldraw[draw=white!80,fill=gray!20] (6.5,1) circle (1.5);
		\draw[red,rotate around={60:(6.5,1)}]  (8,1) arc (0:240:1.5);
		\draw[densely dashed,blue,rotate around={-60:(6.5,1)}](8,1) arc (0:120:1.5);
		\draw [red] (1,0)  arc (-89
		:-88:1);
\draw [red] (1,0)  arc (-89
		:-88:1);
		\draw[black] (1,0) node[below]{$2\beta$};
		\draw [line width =0.5pt,blue] (3.5,0.5)  arc (-92
		:-91:0.5);
		\draw[black] (3.5,0.5) node[below]{$\beta$};
		\draw[black] (6.5,-0.5) node[below]{$3\beta$};
		\draw [line width =0.5pt,densely dashed,rotate around={60:(6.5,1)}] (6.5,1)--(8,1);
		\draw [line width =0.5pt,densely dashed,rotate around={-60:(6.5,1)}]  (6.5,1)--(8,1);
		\draw[black] (6.5,1) node[left]{$\rho_1\rho_2$};
		\draw[line width =0.3pt,rotate around={-60:(6.5,1)}] (6.65,1) arc (0:120:0.15);
		\draw[black] (6.5,1) node[right]{$\frac{2\pi}{3}$};
	\end{tikzpicture}
	\caption{The schematic diagram about the replica trick between thermal states. Here we consider $\rho_1\propto e^{-2\beta H}$ and $\rho_2\propto e^{-\beta H}$. Then, the inverse temperature of replicate thermal state is $3\beta$.}
	\label{p1}
\end{figure}
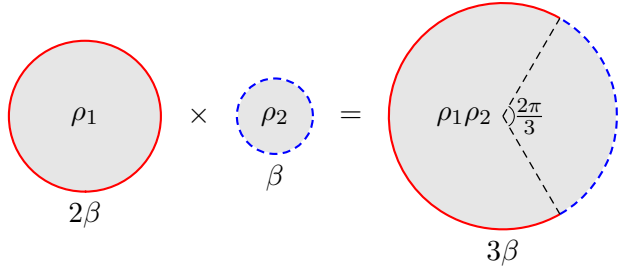

Note that for general two different density matrix $\rho_{1,2}$, we can not guarantee their product $\rho_1\rho_2$ is hermitian. Then, after normalized, the $\rho_1\rho_2/\mathrm{tr}(\rho_1\rho_2)$ is not a density matrix unless satisfying the condition $[\rho_1,\rho_2]=0$. However, for thermal state, $\rho_{1,2}$ commute with each other since they own the same Hamiltonian $H$. The main argument here is that $\rho_1\rho_2$ can be regarded as an \textit{un-normalized} density matrix of a new thermal state with the inverse temperature $\tilde{\beta}=\beta_1+\beta_2$, as depicted in Fig.~\ref{p1}. Note that $\rho_1\rho_2$ is not properly normalized.

\begin{figure}[H]
	\centering
	\begin{tikzpicture}[thick,scale=0.6]
		\draw[rotate =-90] (-2,6) parabola bend (0,0) (2,6);
		\filldraw[draw=white!80,fill=gray!20](7,0)  arc (0:360:1 and 2);
		\draw[red] (6,2)  arc (90:270:1 and 2);
		\draw[densely dashed,blue](6,2) arc (90:-90:1 and 2);
		\draw [line width =1.1pt,loosely dotted] (6,2.3)--(6,1.7);
		\draw [line width =1.1pt,loosely dotted] (6,-2.3)--(6,-1.7);
		\draw [red] (5,0)  arc (180
		:181:1 and 2);
		\draw[black] (5,0) node[left] {$\beta_1$};
		\draw [blue] (7,0.15)  arc (0:1:1 and 2);
		\draw[black] (7,0) node[right] {$\beta_2$};
		\draw[black] (0,0) node[left] {$r_0/2$};
		\filldraw (0,0) circle (0.05);
	\end{tikzpicture}
	\begin{center}
		\begin{tikzpicture}[thick,scale=0.6]
			\draw[rotate =-90] (-2,6) parabola bend (0,0) (2,6);
			\filldraw[draw=white!80,fill=gray!20](7,0)  arc (0:360:1 and 2);
			\draw[red] (6,2)  arc (90:180:1 and 2);
			\draw[densely dashed,blue](5,0) arc (180:270:1 and 2);
			\draw[red] (6,-2)  arc (-90:0:1 and 2);
			\draw[densely dashed,blue](7,0) arc (0:90:1 and 2);
			\draw [line width =1.1pt,loosely dotted] (6,2.4)--(6,1.6);
			\draw [line width =1.1pt,loosely dotted] (6,-2.4)--(6,-1.6);
			\draw [line width =1.1pt,loosely dotted] (4.6,0)--(5.4,0);
			\draw [line width =1.1pt,loosely dotted] (6.6,0)--(7.4,0);
			\draw[black] (6,0) node {$\mathcal{M}_n$};
			\draw[black] (3,0) node {$\mathcal{B}_n$};
			\draw[black] (0,0) node[left] {$\tilde{r}_h~$};
			\filldraw (0,0) circle (0.05);
		\end{tikzpicture}	
	\end{center}
	\caption{The gravity dual of the replicate system by multiplying $n$ thermal states. For general replicate thermal state $\rho_1\rho_2\cdots\rho_n$, its gravity dual is a Schwarzschild-AdS black hole with horizon located at $\tilde{r}_h=1/(r_1^{-1}+r_2^{-1}+\cdots+r_n^{-1})$.}
	\label{p2}
\end{figure}
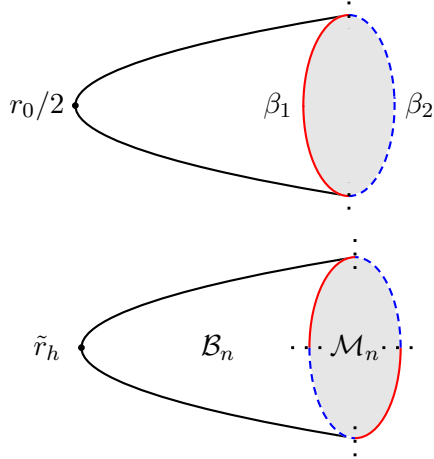

From the bulk perspective, the gravity dual of $\rho_1\rho_2$ would be a Schwarzschild-AdS black hole with new inverse temperature $\tilde{\beta}$. See appendix~\ref{appendix A} for the convention we use here.
According to the relation~\eqref{tem} between temperature $T$ and horizon radius $r_h$, this new black hole's horizon radius $r_h$ can be in principle solved. For convenience, this paper will focus on planar black hole (black brane). For planar Schwarzschild-AdS black hole denoted by $k=0$,
\begin{equation}
	r_h=\frac{4\pi L^2}{d\beta}\,,\beta=\frac{1}{T}\,.
\end{equation}
Then we can get the new horizon radius $\tilde{r}_h$ through $r_{1,2}$, that satisfies $r_{1,2}=4\pi L^2/d\beta_{1,2}$
\begin{equation}
	\tilde{r}_h=\frac{4\pi L^2}{d\tilde{\beta}}=\frac{4\pi L^2}{d(\beta_1+\beta_2)}=\frac{r_1r_2}{r_1+r_2}\,.
\end{equation}
Based on the above calculation, we conclude that the gravity dual of $\rho_1\rho_2$ is Schwarzschild-AdS black hole with smaller horizon radius $\tilde{r}_h$. The product between thermal states gives us more ``smaller'' black hole as depicted in Fig.~\ref{p2}.

After a long journey, we are ready to show how to calculate $\text{tr}(\rho_1\rho_2)$ in Einstein-Hilbert gravity. From the basic holographic dictionary in Ref.~\cite{Maldacena:2001kr}, we have the relation
\begin{equation}
	Z_{\text{CFT}}[\beta]=Z_{\text{bulk}}[\beta]\,,
\end{equation}
that is a most simplest nontrival example of relation~\eqref{dual}. Recalling the relation~\eqref{sad}, here we have
\begin{equation}\label{dic}
	Z[\beta]=\text{tr}e^{-\beta H}=e^{-I[\beta]}\,.
\end{equation}
Ref.~\cite{Emparan:1999pm} shows that, the full Euclidean gravitational action $I[\beta]$ in $d+1$ spacetime dimensions should have three contributions
\begin{equation}\label{total}
	I[\beta]=I_{\text{bulk}}+I_{\text{surf}}+I_{\text{c.t}}\,.
\end{equation}
Here
\begin{equation}
	I_{\text{bulk}}+I_{\text{surf}}=-\frac{1}{16\pi G}\int_{\mathcal{B}}d^{d+1}x\sqrt{g}\left[\mathcal{R}+\frac{d(d-1)}{L^2}\right]-\frac{1}{8\pi G}\int_{\mathcal{\partial B}}d^dx\sqrt{h}K\,.
\end{equation}
The first term is the Einstein-Hilbert action with cosmological constant $-d(d-1)/2L^2$. The second term is called Gibbons-Hawking boundary term that guarantees the well-defined variational principle. The counter term $I_{\text{c.t}}$ is added to regulate the infinity from the classical action. See Ref.~\cite{deHaro:2000vlm,Skenderis:2002wp,Balasubramanian:1999re,Bianchi:2001kw} for details. Here we consider Schwarzschild-AdS planar black hole with inverse temperature $\beta$. The regulated action $I[\beta]$ is given by
\begin{equation}
	\begin{aligned}
		I[\beta]&=-\frac{\beta\Omega_{k=0,d-1}}{16\pi GL^2}r_h^d=-\frac{4^{d-2}\pi^{d-1}L^{2d-2}\Omega_{k=0,d-1}}{d^dG}\frac{1}{\beta^{d-1}}=-\frac{\Omega_{k=0,d-1}r_h^{d-1}}{4Gd}
	\end{aligned}
\end{equation}
Under such holographic description and the saddle point approximation, the trace of $\rho_1\rho_2$ can be expressed in geometric quantities
\begin{equation}
	\begin{aligned}
		\text{tr}(\rho_1\rho_2)=\frac{Z(\beta_1+\beta_2)}{Z(\beta_1)Z(\beta_2)}&=\exp\left(I[\beta_1]+I[\beta_2]-I[\beta_1+\beta_2]\right)\\
		&=\exp\left[{\frac{\Omega_{k=0,d-1}(\tilde{r}_h^{d-1}-r_1^{d-1}-r_2^{d-1})}{4Gd}}\right]
	\end{aligned}
\end{equation}

Above calculation can be generalized into more general case such as tr$(\rho_1\rho_2\rho_3\cdots)$.
\begin{equation}
		\begin{aligned}
		\text{tr}(\rho_1\rho_2\rho_3\cdots)&=\frac{Z(\beta_1+\beta_2+\beta_3+\cdots)}{Z(\beta_1)Z(\beta_2)Z(\beta_3)\cdots}=\exp\left(I[\beta_1]+I[\beta_2]+I[\beta_3]\cdots-I[\beta_1+\beta_2+\beta_3\cdots]\right)\\
		&=\exp\left[{\frac{\Omega_{k=0,d-1}(\tilde{r}_h^{d-1}-r_1^{d-1}-r_2^{d-1}-r_3^{d-1}-\cdots)}{4Gd}}\right]
	\end{aligned}
\end{equation}
Here $\tilde{r}_h=1/(r_1^{-1}+r_2^{-1}+r_3^{-1}+\cdots)$. This gives us a formula to compute the trace of $\rho_1\dots\rho_n$ from holography.

To check our result, one can calculate the R\'{e}nyi entropy for thermal state $\rho$ by taking $n$ copies. According to the definition of R\'{e}nyi entropy, we can obtain that
\begin{equation}
	\begin{aligned}
		S_n=\frac{1}{1-n}\ln\frac{\text{tr}\rho^n}{(\text{tr}\rho)^n}&=\frac{1}{n-1}\left(I_{\text{bulk}}[\mathcal{B}_n]-nI_{\text{bulk}}[\mathcal{B}_1]\right)	\\
		&=\frac{\Omega_{k=0,d-1}r_h^{d-1}}{4G}\frac{n^d-1}{d(n^d-n^{d-1})}\,.
	\end{aligned}
\end{equation}
Following the spirit of the replica trick, then we analytically continue $n$ into unit one
\begin{equation}
	S=\lim\limits_{n\rightarrow1}S_n=\frac{\Omega_{k=0,d-1}r_h^{d-1}}{4G}\,.
\end{equation}
We can see that the Bekenstein-Hawking entropy formula can be precisely recovered. This is an check of self-consistence on our method.
Taken the lessons we learned above, we will consider various quantum distances between thermal states in the remaining content. More specifically, we will employ the method of constructing R\'{e}nyi entropy, i.e. \emph{replica trick}, to calculate these quantum distances.
\subsection{Fidelity}
Given two thermal states $\rho$ and $\sigma$ in the same Hilbert space, we now is ready to calculate the fidelity $\text{Fi}(\rho,\sigma)=$tr$\sqrt{\sqrt{\rho}\sigma\sqrt{\rho}}$ in holography. Since the dimension of Hilbert spacetime is infinity in the field theory, the fidelity of two different states will almost vanish and it will be more useful to consider the logarithmic fidelity
$\ln\text{Fi}(\rho,\sigma)$. Since the square root is involved into the trace, we cannot directly compute the fidelity by using integer-order replica shown in above example. To overcome this issue, we now define a generalized $n$-order log-fidelity for positive integer $n$
\begin{equation}
	\ln\text{Fi}_n(\rho,\sigma)=\ln\text{tr}\left(\sqrt{\sigma}\rho\sqrt{\sigma}\right)^n=\ln\text{tr}(\rho\sigma)^n\,.
\end{equation}
Then, we can analytically continue $n$ into real number. If $\ln\text{Fi}_n(\rho,\sigma)$ is analytic for $n>0$, the fidelity can be recovered
\begin{equation}
	\ln\text{Fi}(\rho,\sigma)=\lim_{n\rightarrow1/2}\ln\text{Fi}_n(\rho,\sigma)\,.
\end{equation}
Generalizing the procedure for $\text{tr}\rho_1\rho_2$ into $\text{tr}(\rho_1\rho_2)^n$
\begin{equation}
	\begin{aligned}
		\text{tr}(\rho_1\rho_2)^n=\frac{Z(n\beta_1+n\beta_2)}{\left[Z(\beta_1)Z(\beta_2)\right]^n}&=\exp\left(nI[\beta_1]+nI[\beta_2]-I[n\beta_1+n\beta_2]\right)\\
		&=\exp\left[{\frac{\Omega_{k=0,d-1}(\tilde{r}_h^{d-1}-nr_1^{d-1}-nr_2^{d-1})}{4Gd}}\right]
	\end{aligned}
\end{equation}
here $\tilde{r}_h(n)=1/[n(r_1^{-1}+r_2^{-1})]$, that is an inverse proportional function of copies number $n$. As just claimed before, we then analytically continue $n$ into $\frac{1}{2}$ to obtain the fidelity
\begin{equation}
	\ln\text{Fi}(\rho,\sigma)=\lim_{n\rightarrow1/2}\ln\text{Fi}_n(\rho,\sigma)={\frac{\Omega_{k=0,d-1}\left[2\tilde{r}_h^{d-1}-r_1^{d-1}-r_2^{d-1}\right]}{8Gd}}\,.
\end{equation}
here $\tilde{r}_h$ take the value $\tilde{r}_h=2r_1r_2/(r_1+r_2)$. As a consistent check, we can confirm that $F(\rho,\sigma)=1$ if $\beta_1=\beta_2$. 
\subsection{Relative entropy}
Relative entropy is a measure of the distinguishability of
two states. It will not suffer from the ultraviolet divergence that carries no information about the state. Given two density matrices, the relative entropy \added{is} defined as
\begin{equation}\label{relative}
	S(\rho||\sigma)=\text{tr}(\rho\ln\rho)-\text{tr}(\rho\ln\sigma)
	\,.
\end{equation}
Here $\rho$ and $\sigma$ are both properly normalized. The first term on the right hand of equal sign is just the entanglement entropy of $\rho$. In order to compute the second term in holography, we can imitate the construction from R\'{e}nyi entropy to entanglement entropy
\begin{equation}
	\text{tr}(\rho\ln\sigma)=\lim_{n\rightarrow1}\frac{1}{n-1}\ln\text{tr}(\rho\sigma^{n-1})\,.
\end{equation}
Still considering two thermal states $\rho_1$ and $\rho_2$, $\text{tr}(\rho_1\rho_2^{n-1})$ is given by
\begin{equation}
	\begin{aligned}
		\text{tr}(\rho_1\rho_2^{n-1})=\frac{Z(\beta_1+(n-1)\beta_2)}{Z(\beta_1)Z(\beta_2)^{n-1}}&=\exp\left(I[\beta_1]+(n-1)I[\beta_2]-I[\beta_1+(n-1)\beta_2]\right)\\
		&=\exp\left[{\frac{\Omega_{k=0,d-1}(\tilde{r}_h^{d-1}-r_1^{d-1}-(n-1)r_2^{d-1})}{4Gd}}\right]
	\end{aligned}
\end{equation}
here $\tilde{r}_h(n)=4\pi L^2/d\left[\beta_1+(n-1)\beta_2\right]$. The second term in Eq.~\eqref{relative} can be recovered by analytically continue $n$ into $1$
\begin{equation}
	\text{tr}(\rho_1\ln\rho_2)=\lim_{n\rightarrow1}\frac{1}{n-1}\ln\text{tr}(\rho_1\rho_2^{n-1})=-\frac{\Omega_{k=0,d-1}r_2^{d-1}}{4Gd}-\frac{\Omega_{k=0,d-1}r_1^{d-1}}{4Gd}\frac{\beta_2(d-1)}{\beta_1}\,.
\end{equation}
Combining the above result with entanglement entropy of $\rho_1$, the relative entropy between thermal states $\rho_1$ and $\rho_2$ is given by
\begin{equation}
	S(\rho_1||\rho_2)=\text{tr}(\rho_1\ln\rho_1)-\text{tr}(\rho_1\ln\rho_2)=-\frac{\Omega_{k=0,d-1}r_1^{d-1}}{4G}+\frac{\Omega_{k=0,d-1}r_2^{d-1}}{4Gd}+\frac{\Omega_{k=0,d-1}r_1^{d-1}}{4Gd}\frac{\beta_2(d-1)}{\beta_1}\,.
\end{equation}
It should be noted that the relative entropy vanishes when $\rho_1$ equates to $\rho_2$, which is a expected result no matter what method is used here.

\section{Add some matter: Maxwell field as the first step }\label{s3}
In this section, we want to generalize our replica trick into the charged black hole. For boundary field theory, we  consider the grand canonical ensemble instead of the canonical ensemble. Given the conserved charge $Q$, we can do the similar construction for the matrix density of the states
\begin{equation}\label{charged}
	\rho=\frac{e^{-\beta(H-\mu Q)}}{Z(\beta,\mu)}\,.
\end{equation}
Here $\mu$ is the chemical potential. The gravity dual of such state~\eqref{charged} is charged black hole in AdS spacetime. Since charge $Q$ is conserved, it should satisfy $[Q,H]=0$. By virtue of such commutativity, we will find that the density matrices of different temperatures and charges are also commutative to each others. Thus we can generalize our above replica trick from neutral system into the charged one. Considering states $\rho_1, \rho_2$
\begin{equation}\label{cdefinition}
	\rho_1=\frac{e^{-\beta_1(H-\mu_1 Q)}}{Z(\beta_1,\mu_1)}\,,\rho_2=\frac{e^{-\beta_2(H-\mu_2 Q)}}{Z(\beta_2,\mu_2)}
\end{equation}
Following the spirit of the previous section, we reconsider $\rho_1\rho_2$ as a new state
\begin{equation}\label{cha}
	\begin{split}
	\rho_1\rho_2=\frac{e^{-\tilde{\beta}(H-\tilde{\mu} Q)}}{Z(\beta_1,\mu_1)Z(\beta_2,\mu_2)}\\
	\tilde{\beta}=\beta_1+\beta_2,~\tilde{\mu}=\frac{\mu_1\beta_1+\mu_2\beta_2}{\beta_1+\beta_2}\,.
	\end{split}
\end{equation}
We can see that the replicate state $\rho_1\rho_2$ have the same construction with the original two states. In holography, this claim means the replicate state $\rho_1\rho_2$ prepared by~\eqref{cdefinition} is still dual to AdS-RN black hole with new parameters ${\tilde{\beta}, \tilde{\mu}}$ defined by~\eqref{cha}.

In gravity side, considering the Einstein-Maxwell theory, the total action~\eqref{total} need to add the contribution from the Maxwell field
\begin{equation}\label{maxwell}
	I_{\text{Maxwell}}=-\frac{1}{16\pi G}\int_{\mathcal{B}}d^{d+1}x\sqrt{g}F_{\mu\nu}F^{\mu\nu}\,.
\end{equation}
If we just consider the RN black hole, the form of the metric coincide with the neutral one~\eqref{metric} but with
\begin{equation}
	f(r)=\frac{r^2}{L^2}+k-\frac{f_0}{r^{d-2}}+\frac{2(d-2)\mu^2r_h^{2(d-2)}}{(d-1)r^{2(d-2)}}\,.
\end{equation}
Here $\mu$ is the chemical potential of the black hole and $r_h$ is the largest root of $f(r)=0$. For RN black hole, the Maxwell field strength tensor has a form
\begin{equation}
	F_{\mu\nu}=-\frac{Q}{r^{d-1}}(d\tau)_\mu\wedge(dr)_\nu,~Q=(d-2)\mu r_h^{d-2}\,.
\end{equation}
Here $Q$ is the total charge of RN black hole. Notice that $F_{\mu\nu}F^{\mu\nu}=2Q^2/r^{2(d-1)}$, the direct calculation for $I_{\text{Maxwell}}$ reads
\begin{equation}
	\begin{aligned}
		I_{\text{Maxwell}}&=-\frac{1}{16\pi G}\int_{\mathcal{B}}d^{d+1}x\sqrt{g}F_{\mu\nu}F^{\mu\nu}\\
		&=-\frac{\beta\Omega_{k, d-1}Q^2r_h^{2-d}}{8\pi G(d-2)}\\
		&=-\frac{(d-2)\beta\Omega_{k, d-1}\mu^2r_h^{d-2}}{8\pi G}
	\end{aligned}
\end{equation}
$\left\{r_h, \beta, \mu\right\}$ is not independent of each other. For $k=0$, we have a relation among $\left\{r_h, \beta, \mu\right\}$:
\begin{equation}\label{horizonRN}
	r_h=\frac{\sqrt{2(d-1)L^2\left[\beta^2d(d-2)^2\mu^2+2\pi^2(d-1)L^2\right]}+2\pi(d-1)L^2}{d(d-1)\beta}
\end{equation}
According to the replica trick~\eqref{cha}, we should consider the gravity dual of replicate state $\rho_1\rho_2$ as a RN-AdS black hole with thermodynamic parameters $\left\{\tilde{\beta},\tilde{\mu}\right\}$. By employing the holographic dictionary
$$
		Z[\beta,\mu]=\text{tr}e^{-\beta (H-\mu Q)}=e^{-I[\beta,\mu]}\,,
$$
we can use the on shell action to express $\text{tr}(\rho_1\rho_2)$
\begin{equation}
	\text{tr}(\rho_1\rho_2)=\frac{Z(\tilde{\beta},\tilde{\mu})}{Z(\beta_1,\mu_1)Z(\beta_2,\mu_2)}=\exp\left(I[\beta_1,\mu_1]+I[\beta_2,\mu_2]-I[\tilde{\beta},\tilde{\mu}]\right)\,.
\end{equation}

For instance, if we set $\left\{\beta_1=\beta_2=\beta\,, \mu_1=\mu_2=\mu\right\}$, the parameters of replicate RN black hole are
\begin{equation}
	\tilde{\beta}=\beta_1+\beta_2=2\beta,~\tilde{\mu}=\frac{\mu_1\beta_1+\mu_2\beta_2}{\beta_1+\beta_2}=\mu\,.
\end{equation}
In general the action $Z(\beta,\mu)$ will be a complicated nonlinear function of $\beta $ and $\mu$. For simplicity of calculation,  we can ignore the back reaction to bulk geometry from the Maxwell field if $\mu\ll1$. In this sense,
\begin{equation}
	r_h\approx\frac{4\pi L^2}{d\beta}\,.
\end{equation}
Then we obtain that
\begin{equation}
	\begin{aligned}
		I[\beta,\mu]&=-\frac{\Omega_{k=0,d-1}r_h^{d-1}}{4Gd}-\frac{(d-2)\Omega_{k=0,d-1}\mu^2r_h^{d-3}L^2}{2Gd}+\mathcal{O}(\mu^4)\\
		I[2\beta,\mu]&=-\frac{\Omega_{k=0,d-1}\tilde{r}_h^{d-1}}{4Gd}-\frac{(d-2)\Omega_{k=0,d-1}\mu^2\tilde{r}_h^{d-3}L^2}{2Gd}+\mathcal{O}(\mu^4)
	\end{aligned}
\end{equation}
Using the fact that $\tilde{r}_h$ is the half of $r_h$ and neglect the terms smaller than $\mathcal{O}(\mu^2)$, it is straight to calculate $\text{tr}(\rho_1\rho_2)$
\begin{equation}
	\begin{aligned}
		\text{tr}(\rho_1\rho_2)&=\exp\left(2I[\beta,\mu]-I[2\beta,\mu]\right)\\
		&=\exp\left[{\frac{\Omega_{k=0,d-1}r_h^{d-1}(1/2^{d-1}-2)}{4Gd}}+\frac{(d-2)\Omega_{k=0,d-1}\mu^2L^2r_h^{d-3}(1/2^{d-3}-2)}{2Gd}\right]
	\end{aligned}
\end{equation}
Finally, we generalize our result into $n$ copies and calculate the R\'{e}nyi entropy
\begin{equation}
	S_n=\frac{\Omega_{k=0,d-1}r_h^{d-1}}{4Gd}\frac{n^d-1}{n^d-n^{d-1}}+\frac{(d-2)\Omega_{k=0,d-1}\mu^2L^2r_h^{d-3}}{2Gd}\frac{n^{d-2}-1}{n^{d-2}-n^{d-3}}\,.
\end{equation}
By taking the limit $n\rightarrow1$, the Von Neumann entropy can be recovered
\begin{equation}\label{rns1}
	S=\lim\limits_{n\rightarrow1}S_n=\frac{\Omega_{k=0,d-1}r_h^{d-1}}{4G}+\frac{(d-2)^2\Omega_{k=0,d-1}\mu^2L^2r_h^{d-3}}{2Gd}\,.
\end{equation}

One can also get the above result according to Bekenstein-Hawking formula. Since we consider the chemical potential $\mu$ as a infinitesimal, the horizon~\eqref{horizonRN} of RN-AdS black hole denoted by $r_{\text{RN}}$ can be separated into two parts
\begin{equation}
	r_{\text{RN}}=r_h+\frac{2(d-2)^2L^2\mu^2}{d(d-1)r_h}\,.
\end{equation}
Here $r_h$ is the largest root of $f(r)=0~\text{for}~ \mu=0$. Following the Bekenstein-Hawking  formula, the entropy is described by the area of horizon. Up to the order $\mathcal{O}(\mu^2)$ we obtain
\begin{equation}\label{bks2}
	\begin{aligned}
		S&=\frac{\Omega_{k=0,d-1}r_{\text{RN}}^{d-1}}{4G}\\
		&=\frac{\Omega_{k=0,d-1}r_h^{d-1}}{4G}+\frac{(d-2)^2\Omega_{k=0,d-1}\mu^2L^2r_h^{d-3}}{2Gd}\,.
	\end{aligned}
\end{equation}
We see that two formulas \eqref{rns1} and \eqref{bks2} are same, which is what we expect and shows the consistence of our methods.

\section{Fidelity between states excited by scalar operator}\label{s4}
In this section, we consider how to compute the quantum distance between states that are excited by scalar operator. In holography, the scalar operator $\mathcal{O}_\phi$ in boundary CFT is dual to a bulk scalar field  $\phi$. Considering a $(d+1)$-dimensional Einstein-scalar theory, such theory's action is
\begin{equation}\label{to}
	I=-\frac{1}{16\pi G}\int_{\mathcal{B}}d^{d+1}x\sqrt{g_E}\left[\mathcal{R}+\frac{d(d-1)}{L^2}-\frac{1}{2}\nabla_\mu\phi\nabla^\mu\phi-\frac{1}{2}m^2\phi^2\right]\,.
\end{equation}
The parameter $m$ is the mass of the scalar field. Different from the Minkowski spacetime, $m^2$ is negative in AdS spacetime. Despite this, $m^2$ can not be too negative but above the  Breitenlohner-Freedman bound $m_{\text{BF}}^2\equiv-d^2/4L^2$. For dual boundary CFT, the corresponding conformal dimension $\Delta_{\mathcal{O}}$ for scalar operator is $(d\pm\sqrt{d^2+4m^2L^2})/2$. The ``$+$'' corresponding to standard quantization and ``$-$'' stands for alternative quantization. We can get the equations of motion following the action~\eqref{to}
\begin{equation}
	\begin{aligned}
		\mathcal{R}_{\mu\nu}-\frac{1}{2}\mathcal{R}g_{\mu\nu}-\frac{d(d-1)}{2L^2}g_{\mu\nu}&=-\frac{1}{4}\left(\nabla_\rho\phi\nabla^{\rho}\phi+m^2\phi^2\right)
		\\
		\nabla_\mu\nabla^\mu\phi-m^2\phi&=0\,.
	\end{aligned}
\end{equation}
In most general cases, we should solve this coupled gravitational system since the backreaction of the scalar field can not be neglected. In principle, given the proper boundary conditions, these equations can be solved analytically or numerically. We consider the thermal states in the canonical ensemble. However, we can neglect the backreaction of the scalar field in the bulk, if the change of the total energy $\Delta E$ induced from the scalar operator $\mathcal{O}_\phi$ is much smaller than the mass of black hole. This is the so-called \emph{probe limit}. In the probe limit, the gravity and scalar field will decouple. That means in contact with a heat bath at $\beta$, the bulk geometry remains to be Schwarzschild-AdS black hole. Then, we just analyze the
behavior of scalar field in this fixed black hole background. What's more, we only need to explore the additional contributions from the scalar field to holographic quantum distance. In this paper, we focus on the fidelity between states excited by scalar operator since it is easy to obtain. The main reason why it is tricky to calculate other quantum distances will be explained soon.

\subsection{Set up}
In this section, we adopt the Poincar\'{e} coordinate $\left\{z, \tau, x^i\right\}$. Under this gauge, the $n$-copies Euclidean Schwarzschild-AdS black hole's metric reads
\begin{equation}\label{zgauge}
	ds^2=\frac{1}{z^2}\left[f(z)d\tau^2+\frac{dz^2}{f(z)}+\sum_{i=1}^{d-1} d x_{i}^{2}\right],\quad f(z)=\frac{1}{L^2}-(2nz)^d f_0\,.
\end{equation}
Here the factor $2n$ is induced from the definition of fidelity where $\text{tr}(\rho\sigma)^n$ is involved. Based on our previous discussion, for two states $\rho$ and $\sigma$ with same inverse temperature $\beta$, the dual black hole to compute $\text{tr}(\rho\sigma)^n$ will have inverse temperature $2n\beta$. Then, we write the equation of motion for scalar field in this background:
\begin{equation}
	\frac{1}{\sqrt{g_E}}\frac{\partial}{\partial x^\mu}\left(\sqrt{g_E}g^{\mu\nu}\frac{\partial}{\partial x^\nu}\right)\phi-m^2\phi=0,\quad g_E=\frac{\Omega_{0,d-1}}{z^{2(d+1)}}\,.
\end{equation}
More specifically,
\begin{equation}\label{eofphi}
\frac{z^2\ddot{\phi}}{f}+z^2f\phi''+\left[z^2f'-(d-1)zf\right]\phi'-m^2\phi=0\,,
\end{equation}
where a prime/dot denotes the derivative with respect to $z$ or $\tau$, respectively. In general case, the Eq.~\eqref{eofphi} does not have analytical solutions so we have to solve this equation numerically. Here we assume the boundary can keep the asymptotic AdS behavior such that
\begin{equation}\label{b}
	\phi(\tau,z)\rightarrow\tilde{\phi}_s(\tau)z^{d-\Delta}+\tilde{\phi}_v(\tau)z^\Delta+\dots\,,~~~\Delta=\frac{d+\sqrt{d^2+4m^2L^2}}{2}\,.
\end{equation}
Other requirement is that $\phi$ should be continuous near the horizon $z_h$. Here $\tilde{\phi}_s(\tau)$ and $\tilde{\phi}_v(\tau)$ are independent of $z$. In holography, both $\tilde{\phi}_s(\tau)$ and $\tilde{\phi}_v(\tau)$
can be considered as a source if the mass parameter satisfies $m_{\text{BF}}^2<m^2<m_{\text{BF}}^2+1/L^2$. It needs to note that $\tilde{\phi}_s$ and $\tilde{\phi}_v$ will be both nonzero if the scalar field is not zero in the bulk. To keep the asymptotically AdS structure and make the probe limit self-consistent we have to require $\phi^2\rightarrow0$ as $z\rightarrow0$. This requires that we have to set $d>\Delta$ and so we have to set:
\begin{equation}\label{negmass2}
  m_{\text{BF}}^2<m^2<0\,.
\end{equation}
The selection of source corresponds different dual field theories, see Ref.~\cite{Klebanov:1999tb}. In gravity side, this selection determines how to  set the boundary condition for Eq.~\eqref{eofphi}. For instance, if considering $\tilde{\phi}_s(\tau)$ as source, we must fix $\tilde{\phi}_s(\tau)$ at $z\rightarrow0$ correspondingly. This is called \emph{standard quantization scheme} for the scalar field. While in this paper, we employ the \emph{alternative quantization scheme} that considers $\tilde{\phi}_v(\tau)$ as the source. The reason why we choose ``non-standard'' quantization scheme will be explicitly explained. It should be noted that the scalar field obeys second order equation of motion~\eqref{eofphi}, but we only give one boundary condition such as fixing $\tilde{\phi}_v(\tau)$ in alternative quantization scheme. However, we must discard one of the independent solutions since this solution blows up at the horizon $z_h$. Hence, the regularity of Euclidean spacetime makes the bulk scalar field uniquely specified.

 The next step is to solve Eq.~\eqref{eofphi} since the final result for action~\eqref{to} is dependent of $\tilde{\phi}_v(\tau)$ and $\tilde{\phi}_s(\tau)$. Then we can assume the general solution of $\phi(\tau,z)$ by the method of separation of variables
\begin{equation}
	\phi(\tau,z)=\sum_{j=0}^{\infty}R_j(z)T_j(\tau)\,.
\end{equation}
Substituting this into Eq.~\eqref{eofphi}, we can get the asymptotic behavior of general solution for scalar field
\begin{equation}
	\begin{aligned}
		\lim_{z\rightarrow0}\phi(\tau,z)
		&=\lim_{z\rightarrow0}\sum_{j=0}^{\infty}T_j(\tau)R_j(z) \\
		T_0(\tau)&=A_0\,,T_j(\tau)=A_j\cos w_j\tau+B_j\sin w_j \tau\,,\\
		\lim_{z\rightarrow0}R_j(z)&=\phi_s^{(j)}z^{d-\Delta}(1+\cdots)+\phi_v^{(j)}z^{\Delta}(1+\cdots)\,.
	\end{aligned}
\end{equation}
and the radial equation
\begin{equation}\label{eomode1}
	\begin{split}
		R_j''+\left(\frac{f'}{f}-\frac{d-1}{z}\right)R_j'-\left(\frac{m^2}{z^2f}+\frac{w_j^2}{f^2}\right)R_j=0\\
		f(z)=\frac{1}{L^2}-(2nz)^d f_0\,,~~j=0,1,2,\dots
	\end{split}
\end{equation}
Here $\phi_{s,v}^{(j)}$ is denoted by the asymptotic behavior of $R_j (z)$. The frequency $w_j$ take some discrete values given by
\begin{equation}
	w_j=\frac{2\pi j}{\tilde{\tau}}\,,\quad T(\tau)=T(\tau+\tilde{\tau})
\end{equation}
where $\tilde{\tau}$ is the period in boundary condition \eqref{b}. According to Fourier transformation, coefficients $A_0, A_n, B_n$ are determined by
\begin{equation}
	\begin{aligned}
		A_0\phi_v^{(0)}&=\frac{1}{\tilde{\tau}}\int_{0}^{\tilde{\tau}}\tilde{\phi}_v(\tau)\mathrm{d}\tau \\
		A_j\phi_v^{(j)}&=\frac{2}{\tilde{\tau}}\int_{0}^{\tilde{\tau}}\tilde{\phi}_v(\tau)\cos w_j\tau\mathrm{d}\tau \\
		B_j\phi_v^{(j)}&=\frac{2}{\tilde{\tau}}\int_{0}^{\tilde{\tau}}\tilde{\phi}_v(\tau)\sin w_j\tau\mathrm{d}\tau
	\end{aligned}
\end{equation}
\subsection{Fidelity in the probe limit}
In following we will first focus on \textit{alternative quantization}, i.e. treating the coefficient $\phi_v$ as the source term. As previously defined, given two density matrices $\rho$ and $\sigma$ that are both excited by scalar operator $\mathcal{O_\phi}$ with different values of source $\tilde{\phi}_v$, a generalized $n$-order log-fidelity for positive integer $n$ is
\begin{equation}
	\ln\text{Fi}_n(\rho,\sigma)=\ln\text{tr}\left(\sqrt{\sigma}\rho\sqrt{\sigma}\right)^n=\ln\text{tr}(\rho\sigma)^n\,.
\end{equation}
Then, we can analytically continue $n$ into $1/2$ to recover the fidelity
\begin{equation}
	\ln\text{Fi}(\rho,\sigma)=\lim_{n\rightarrow1/2}\ln\text{Fi}_n(\rho,\sigma)\,.
\end{equation}

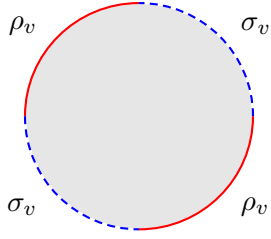
\begin{figure}[H]
	\centering
		\begin{tikzpicture}[thick, scale=1.5]
			\filldraw[draw=white!80,fill=gray!20] (1,1) circle (1);
			\draw[red] (1,2) arc (90:180:1);
			\draw[densely dashed,blue](0,1) arc (180:270:1);
			\draw[red](1,0) arc (270:360:1);
			\draw[densely dashed,blue](2,1) arc (0:90:1);
			\draw[black] (0.2,1.8) node[left] {$\rho_v$};
			\draw[black] (0.2,0.2) node[left] {$\sigma_v$};
			\draw[black] (1.8,0.2) node[right] {$\rho_v$};
			\draw[black] (1.8,1.8) node[right] {$\sigma_v$};
		\end{tikzpicture}%
		\caption{The schematic diagram about the corresponding bulk geometry with the boundary field configuration $\rho\sigma\rho\sigma$. Here the red line labels the distribution of state $\rho$ and blue dashed line labels the state $\sigma$. Here subscript $v$ follows from treating the coefficient $\phi_v$ as the source term under alternative quantization. As setting $\beta_1=\beta_2=\beta$, we can see that the period $\tilde{\tau}=2\beta$.}
		\label{n2}
\end{figure}

Next we construct the corresponding distribution $\tilde{\phi}_v(\tau)$ for $\text{tr}(\rho\sigma)^n$ in the bulk, see Fig.~\eqref{n2} for illustration of $\text{tr}(\rho\sigma)^2$. In the bulk gravity side, the boundary condition for scalar field $\phi$ will be dependent of euclidean time
\begin{equation}\label{boundary}
	\tilde{\phi}_v(\tau)=\left\{\begin{aligned}
		\rho_v\quad 0<\tau\leq\beta \\
		\sigma_v\quad\beta<\tau\leq2\beta \\
		\rho_v\quad 2\beta<\tau\leq3\beta \\
		\dots\\
		\rho_v\quad 2(n-2)\beta<\tau\leq 2(n-1)\beta \\
		\sigma_v\quad 2(n-1)\beta<\tau\leq 2n\beta \\
	\end{aligned}
	\right\}\,.
\end{equation}
Here $\rho_v$ and $\sigma_v$ are constants. We already set $\beta_1=\beta_2=\beta$. To simplify the discussion, in this paper we just set $\sigma_v=0$.
In principle, we need to know the value of bulk scalar field in every spacetime point. Since the total action~\eqref{to} is a spacetime volume integral. However, according to Gauss theorem and using the equation of motion, the on-shell action for scalar field can be expressed by the boundary term
\begin{equation}
	\begin{aligned}
		I_m
		&=\frac{1}{16\pi G}\int_{\mathcal{B}_n}\mathrm{d}^{d+1}x\sqrt{g}\left[\frac{1}{2}\nabla_\mu\left(\phi\nabla^\mu\phi\right)-\frac{1}{2}\phi\nabla_\mu\nabla^\mu\phi+\frac{1}{2}m^2\phi^2\right]\\
		&=\frac{1}{16\pi G}\int_{\mathcal{B}_n}\mathrm{d}^{d+1}x\sqrt{g}\left[\frac{1}{2}\nabla_\mu\left(\phi\nabla^\mu\phi\right)\right]\\
		&=\frac{1}{16\pi G}\int_{\partial\mathcal{B}_n}\mathrm{d}^{d}x\sqrt{h}\left(\frac{1}{2}n_\mu\phi \nabla^\mu\phi\right)\,.
	\end{aligned}
\end{equation}
Here $n^\mu$ is the outward point normal vector of AdS boundary $\partial\mathcal{B}_n$. In Poincar\'{e} coordinate, the only nonzero component is
\begin{equation}
	n^r=-z\sqrt{f(z)}\,,
\end{equation}
so this boundary integral can reduce to
\begin{equation}
	I_m=-\frac{\Omega_{0,d-1}}{32\pi G}\int\mathrm{d}\tau\left(\frac{f\phi\partial_z \phi}{z^{d-1}}\right) \,.
\end{equation}
Note that the above action is not regulated. In principle, according to spacetime dimension $d$ and conformal dimension $\Delta$, we will correspondingly select different forms of counter term~\cite{deHaro:2000vlm,Skenderis:2002wp,Balasubramanian:1999re,Bianchi:2001kw}. In this paper, we will focus on 4-dimensional Einstein-scalar theory while fixing $m^2=-2$. For $\left\{d=3,\Delta=2\right\}$, the corresponding counter term is
\begin{equation}
	I_\text{c.t}=-\frac{1}{16\pi G}\int d^3x\sqrt{h}\left[\phi(n^\mu\partial_\mu\phi)+\frac{1}{2}\phi^2\right]\,.
\end{equation}
Therefore, the regulated scalar field action $\tilde{I}_{\rho\sigma}$ can be easily calculated
\begin{equation}
	\begin{aligned}\label{sumI1}
		\tilde{I}_{\rho\sigma}&=I_m+I_{\text{c.t}}\\
		&=\frac{1}{32\pi G}\int\mathrm{d}^{3}x\left(\phi_s^{(0)}A_0+\sum_{i=1}^{\infty}\phi_s^{(i)}B_i\sin w_i\tau\right)\left(\phi_v^{(0)}A_0+\sum_{j=1}^{\infty}\phi_v^{(j)}B_j\sin w_j\tau\right)\\
		&=\frac{2n\beta\Omega_{0,2}}{32\pi G}\left(\phi_s^{(0)}\phi_v^{(0)}A_0^2+\frac{1}{2}\sum_{j=1}^{\infty}\phi_s^{(j)}\phi_v^{(j)}B_j^2\right)\\
		&=\frac{n\beta\Omega_{0,2}}{16\pi G}\left[\frac{\rho_v^2\phi_s^{(0)}}{4\phi_v^{(0)}}+\sum_{j=1}^{\infty}\frac{(1-\cos j\pi)^2}{j^2\pi^2}\frac{\rho_v^2\phi_s^{(j)}}{2\phi_v^{(j)}}\right]\\
		&\equiv\frac{n\beta\Omega_{0,2}}{16\pi G}\left[\frac{\rho_v^2\phi_s^{(0)}}{4\phi_v^{(0)}}+\Xi[\rho_v,n]\right]\,.
	\end{aligned}
\end{equation}
Here
\begin{equation}\label{defxi1}
\Xi[\rho_v,n]:=\sum_{j=1}^{\infty}\frac{(1-\cos j\pi)^2}{j^2\pi^2}\frac{\rho_v^2\phi_s^{(j)}}{2\phi_v^{(j)}}\,.
\end{equation}
Before we go further, it needs to emphasize that divergency of ``naked'' action arises only when we approaches to UV boundary $z\rightarrow0$. The counterterm then is applied to cancel such divergency. We will see later that, since the boundary is inhomogeneous, there is an other different divergency of on-shell action appears in standard quantization even if we do not take the limit $z=0$. This kind of new divergency cannot be cancelled according to the known holographic renormalization schema~\cite{deHaro:2000vlm,Skenderis:2002wp,Balasubramanian:1999re,Bianchi:2001kw}.

We now return to Eq.~\eqref{sumI1}. Notice that $\Xi$ is dependent on the copies number $n$, since the radial equation is related to $n$, i.e. $f(z)=1/L^2-(2nz)^d f_0$. This property is very different from the usual cases that $\tilde{\phi}_v$ is independent of the Euclidean time $\tau$. If $\Xi[\rho_v,\frac{1}{2}]$ is convergent, then we would analytically continue $n$ into $1/2$. In other words, we need to solve the radial equation~\eqref{eomode1} with $f(z)=1/L^2-z^df_0$. Finally, the log-fidelity reads
\begin{equation}
	\begin{aligned}
	\ln\text{Fi}(\rho,\sigma)&=\lim_{n\rightarrow1/2}\ln\frac{\text{tr}(\rho\sigma)^n}{(\text{tr}\rho)^n(\text{tr}\sigma)^n}\\
	&=\lim_{n\rightarrow1/2}\ln\exp[-\tilde{I}_{\rho\sigma}+n\tilde{I}_\rho]\\
	&=\lim_{n\rightarrow1/2}\frac{n\beta\Omega_{0,2}}{16\pi G}\left[\frac{\rho_v^2\phi_s^{(0)}}{4\phi_v^{(0)}}-\Xi[\rho_v,n]\right]\\
	&=\frac{\beta\Omega_{0,2}}{32\pi G}\left[\frac{\rho_v^2\phi_s^{(0)}}{4\phi_v^{(0)}}-\Xi[\rho_v,\frac{1}{2}]\right]\,.
	\end{aligned}
\end{equation}
Here $\tilde{I}_\rho$ is the regulated action for $\text{tr}\rho$
\begin{equation}\label{eqIrho1}
	\begin{aligned}
		\tilde{I}_\rho&=\frac{1}{32\pi G}\int\mathrm{d}^{3}x\left[\phi_s^{(0)}\phi_v^{(0)}A_0^2\right]\\
		&=\frac{1}{32\pi G}\int\mathrm{d}^{3}x\left[\frac{\rho_v^2\phi_s^{(0)}}{\phi_v^{(0)}}\right]\\
		&=\frac{\beta\Omega_{0,2}}{32\pi G}\frac{\rho_v^2\phi_s^{(0)}}{\phi_v^{(0)}}\,.
	\end{aligned}
\end{equation}
Note again that Eq.~\eqref{eomode1} can be solved for arbitrary real number $n$. Thus, we can numerically compute the value of $\Xi[\rho_v,n]$ in the Eq.~\eqref{defxi1} and then obtain the fidelity numerically.

\subsection{Numerical result}\label{num}
After the previous derivation for fidelity, we then show the corresponding numerical process and result. The only remaining unsolved equation is the radial equation
\begin{equation}\label{radial}
	\begin{split}
		R_j''+\left(\frac{f'}{f}-\frac{d-1}{z}\right)R_j'-\left(\frac{m^2}{z^2f}+\frac{w_j^2}{f^2}\right)R_j=0\\
		f(z)=\frac{1}{L^2}-(2nz)^d f_0\,,~~j=0,1,2,\dots
	\end{split}
\end{equation}
We need to get the value of $\phi_s^{(j)}/\phi_v^{(j)}$ in $\Xi[\rho_v,n]$ for every frequency $w_j$. For $j=0$, the frequency is vanished, i.e. $w_0=0$. The general solution for $R_0$ is just the static case
\begin{equation}
	\begin{aligned}
		R_0=&\phi_s^{(0)}z^{d-\Delta}{_2F_1}\left(1-\frac{\Delta}{d}\,,1-\frac{\Delta}{d}\,;2-\frac{2\Delta}{d}\,;f_0L^2n^dz^d\right)\\
		+&\phi_v^{(0)}z^{\Delta}{_2F_1}\left(\frac{\Delta}{d}\,,\frac{\Delta}{d}\,;\frac{2\Delta}{d}\,;f_0L^2n^dz^d\right)\,.
	\end{aligned}
\end{equation}
Continuity condition requires that $R_0$ should be finite at horizon $z_h$ which satisfies
\begin{equation}
	\frac{1}{L^2}-f_0n^dz_h^d=0\,.
\end{equation}
By virtue of
\begin{equation}\label{gamma}
	{_2F_1}(a\,,b\,;c\,;1)=\frac{\Gamma(c)\Gamma(c-a-b)}{\Gamma(c-a)\Gamma(c-b)}\,,
\end{equation}
$R_0(z_h)$ can be expressed as
\begin{equation}
		R_0(z_h)=\phi_s^{(0)}z_h^{d-\Delta}\frac{\Gamma(2-\frac{2\Delta}{d})\Gamma(0)}{\Gamma(1-\frac{\Delta}{d})\Gamma(1-\frac{\Delta}{d})}
		+\phi_v^{(0)}z_h^{\Delta}\frac{\Gamma(\frac{2\Delta}{d})\Gamma(0)}{\Gamma(\frac{\Delta}{d})\Gamma(\frac{\Delta}{d})}\,.
\end{equation}
Notice that $R_0(z_h)$ is singular since $\Gamma(0)$ is infinite. We need to eliminate this singularity. Extract $\Gamma(0)$ outside the square bracket
\begin{equation}
	R_0(z_h)=\Gamma(0)\left[\phi_s^{(0)}z_h^{d-\Delta}\frac{\Gamma(2-\frac{2\Delta}{d})}{\Gamma(1-\frac{\Delta}{d})\Gamma(1-\frac{\Delta}{d})}
	+\phi_v^{(0)}z_h^{\Delta}\frac{\Gamma(\frac{2\Delta}{d})}{\Gamma(\frac{\Delta}{d})\Gamma(\frac{\Delta}{d})}\right]\,.
\end{equation}
Therefore, the square bracket must vanish and the ratio $\phi_s^{(0)}/\phi_v^{(0)}$ is
\begin{equation}
	\frac{\phi_s^{(0)}}{\phi_v^{(0)}}=-\frac{\Gamma(\frac{2\Delta}{d})\Gamma^2(1-\frac{\Delta}{d})}{\Gamma(2-\frac{2\Delta}{d})\Gamma^2(\frac{\Delta}{d})}z_h^{2\Delta-d}=\frac{\pi d}{2\Delta-d}\frac{\sin{\frac{2\pi\Delta}{d}}\Gamma^2(\frac{2\Delta}{d})}{\sin^2{\frac{\pi\Delta}{d}}\Gamma^4(\frac{\Delta}{d})}z_h^{2\Delta-d}\,.
\end{equation}

For $j=1,2,3,\dots$, Eq.~\eqref{radial} does not have analytical solutions so we have to solve the radial equation numerically. Notice that $\Xi[\rho_v,n]$ is the sum of an infinite series. Although we cannot analytically get the explicit expression of $\phi_s^{(j)}/\phi_v^{(j)}$, the convergence or not of $\Xi[\rho_v,n]$ can be obtained from the comparison with the scalar field in thermal AdS spacetime given $f_0=0$ in metric~\eqref{zgauge}. Considering the scalar field with the same boundary condition~\eqref{boundary} in thermal AdS spacetime, the corresponding radio $\phi_s^{(j)}/\phi_v^{(j)}$ can be analytically solved:
\begin{equation}
	\left.\frac{\phi_s^{(j)}}{\phi_v^{(j)}}\right|_{\text{thermal AdS}}=\frac{\Gamma(\Delta-\frac{d}{2})}{(w_j/2)^{2\Delta-d}\Gamma(\frac{d}{2}-\Delta)}=-\frac{\sin[\pi(\Delta-\frac{d}{2})](\Delta-\frac{d}{2})\Gamma(\Delta-\frac{d}{2})^2}{\pi (w_j/2)^{2\Delta-d}}\,.
\end{equation}
Notice that in thermal AdS spacetime $\Xi[\rho_v,\frac{1}{2}]_{\text{thermal AdS}}$ is convergent since the general term is proportional to $1/j^3$. To determine whether $\Xi[\rho_v,\frac{1}{2}]$ is convergent, we give a concrete numerical result to check the change of $\delta_j$ with $j$
\begin{equation}
	\delta_j\equiv\frac{\phi^{(j)}_s}{\phi^{(j)}_v}-\left.\frac{\phi_s^{(j)}}{\phi_v^{(j)}}\right|_{\text{thermal AdS}}
\end{equation}
\begin{figure}[H]
	\centering
	\includegraphics[width=0.8\textwidth]{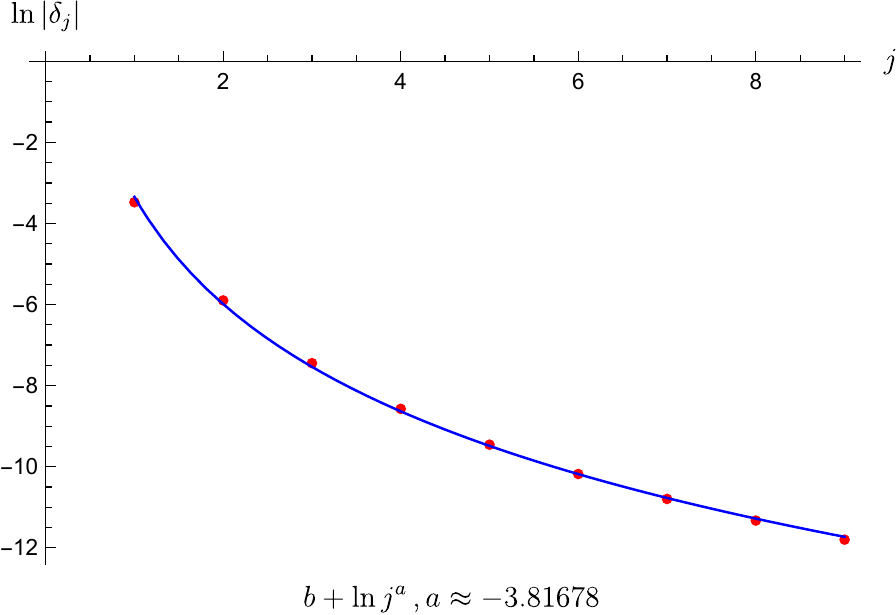}
	\caption{The curve-fitting result of $\delta_j$. Here we use a logarithmic function $b+\ln j^a$ to fit their difference. }
	\label{aa}
\end{figure}
Here we fix the following parameters
$$
m^2=-2,~d=3,~L^2=1,~ f_0=1\,.
$$
From above parameters, we can obtain $\beta=4\pi/3$ and $\Delta=2$. We solve the equation~\eqref{radial} numerically and then read the radio $\phi^{(j)}_s/\phi^{(j)}_v$ from the asymptotic behavior of the radial equation.

According to the fitting result, i.e. Fig.~\eqref{aa}, we conclude that the difference between $\frac{\phi^{(j)}_s}{\phi^{(j)}_v}$ and $\left.\frac{\phi_s^{(j)}}{\phi_v^{(j)}}\right|_{\text{thermal AdS}}$ is $j^a$ order. Since $a\approx-3.81678$, the difference between two series is convergent. It should be noted that the numerical result of $a$ is very close to $-4$. In practice, we can confirm that
\begin{equation}
	\frac{\phi^{(j)}_s}{\phi^{(j)}_v}-\left.\frac{\phi_s^{(j)}}{\phi_v^{(j)}}\right|_{\text{thermal AdS}}=\mathcal{O}(1/w_j^{2\Delta})
\end{equation}
with a analytical approximation method. See appendix~\ref{appendix D} for more details about the solution for radial equation.
Then, according to the comparison test, $\Xi[\rho_v,\frac{1}{2}]$ must be convergent under the alternative quantization scheme since in this scheme $\Xi[\rho_s,\frac{1}{2}]_{\text{thermal AdS}}$ is convergent. The final numerical result reads
\begin{equation}
\begin{aligned}
	\frac{\rho_v^2\phi_s^{(0)}}{4\phi_v^{(0)}}&\approx-1.29053\rho_v^2\\
	\Xi[\rho_v,\frac{1}{2}]&\approx-0.0176151\rho_v^2
\end{aligned}
\end{equation}

\subsection{Comment on standard quantization}
We have shown that the on-shell action of replica spacetime is convergent in alternative quantization. However, under the standard quantization the situation is very different. In this case, we still have Eqs.~\eqref{sumI1}-\eqref{eqIrho1}, but the $\rho_v$ is replaced by $\rho_s$ and $\Xi$ should be replaced by
\begin{equation}\label{defxi13}
\Xi[\rho_s,n]:=\sum_{j=1}^{\infty}\frac{(1-\cos j\pi)^2}{j^2\pi^2}\frac{\rho_s^2\phi_v^{(j)}}{2\phi_s^{(j)}}\,.
\end{equation}
We will find that $\Xi[\rho_s,\frac{1}{2}]_{\text{thermal AdS}}$ is divergent. Specifically, $\Xi[\rho_s,\frac{1}{2}]_{\text{thermal AdS}}$ is a \emph{harmonic series} under the standard quantization since the general term in $\Xi[\rho_s,\frac{1}{2}]_{\text{thermal AdS}}$ becomes
\begin{equation}
	\Xi\sim\sum\frac{\phi^{(j)}_v}{j^2\phi^{(j)}_s}\sim\frac{1}{j}\,.
\end{equation}
The high frequency scalar modes with $w_j\gg1$ will be not suppressed enough and so will play more and more important roles. When it comes to
the computation of action, this effect will cause the infinity of $\Xi[\rho_v,n]$ in the standard quantization scheme. While under alternative quantization scheme, these UV contributions from high frequency modes are suppressed enough. It should be noted that this behavior is quite common when considering inhomogeneous boundary condition for bulk matter. It needs to emphasize that such as divergency appearing in standard quantization cannot be removed by using the usual so called ``holographic renormalization'', since one can realize that this divergency will appear even we move the AdS boundary into finite $z$. We note similar thing was also noticed by Ref.~\cite{Yang:2022ibq}, which finds that generalized susceptibly in terms of momentum integration will be divergent in standard quantization.

One may seek to regulate this infinity in standard quantization scheme. For instance, given a UV cutoff $k$, we seem to be able to extract divergence from the harmonic series
\begin{equation}\label{larhesumk}
	\sum_{j=1}^{k}\frac{1}{j}=\ln k+\gamma+\mathcal{O}(1/k)\,~~k\gg1
\end{equation}
here $\gamma$ is Euler-Mascheroni constant. So the divergence here is logarithmic divergence. When the mode $j$ is high enough, the probe limit and even including the classical gravity approximation will be broken, so we should take the full backreaction and even full UV quantum gravity theory into account. If we believe the theory is UV complete, then ``unknown'' high energy physics should just cancel such a logarithmic divergency. Then we will throw away $\ln k$, which will be interpreted as the contribution from the deformation of spacetime due to these high frequency modes. However, this procedure is only effective for the set of parameters we have selected here. When we fine-tune these parameters, such as $m^2$, the above procedure is broken and we need to regulate such infinity again in this new set of parameters. From the Eq.~\eqref{wj} in appendix~\ref{appendix D} one can see that,
\begin{equation}\label{wjs1}
	\frac{\phi_v^{(j)}}{\phi_s^{(j)}}\sim w_j^{2\Delta-d}\,.
\end{equation}
for large mode $j$. Thus, in standard quantization, we can find from Eq.~\eqref{defxi13} that
\begin{equation}
	\Xi\sim\sum j^{2\Delta-d}\,.
\end{equation}
Since $2\Delta-d=\sqrt{d^2+4m^2L^2}>0$, we see that $\Xi$ will be divergent. It is still an open question for us about how to regularized and renormalize such a divergency systematically. We leave the general regularization under the standard quantization scheme for an important future problem.

 \section{Commutativity of $\rho$ and $\sigma$}\label{sec5}
\begin{figure}[H]
	\centering
	\begin{subfigure}[h]{0.3\linewidth}
		\centering
		\begin{tikzpicture}[thick, scale=1.3]
			\filldraw[draw=white!80,fill=gray!20] (1,1) circle (1);
			\draw[red] (1,2) arc (90:180:1);
			\draw[densely dashed,blue](0,1) arc (180:270:1);
			\draw[red](1,0) arc (270:360:1);
			\draw[densely dashed,blue](2,1) arc (0:90:1);
			\draw[black] (0.2,1.8) node[left] {$\rho$};
			\draw[black] (0.2,0.2) node[left] {$\sigma$};
			\draw[black] (1.8,0.2) node[right] {$\rho$};
			\draw[black] (1.8,1.8) node[right] {$\sigma$};
		\end{tikzpicture}%
		\caption{$\rho\sigma\rho\sigma$} \label{fig:M1}
	\end{subfigure}
	\begin{subfigure}[h]{0.3\linewidth}
		\centering
		\begin{tikzpicture}[thick, scale=1.3]
			\filldraw[draw=white!80,fill=gray!20] (1,1) circle (1);
			\draw[red] (1,2) arc (90:270:1);
			\draw[densely dashed,blue](1,2) arc (90:-90:1);
			\draw[black] (0.2,1.8) node[left] {$\rho$};
			\draw[black] (0.2,0.2) node[left] {$\rho$};
			\draw[black] (1.8,0.2) node[right] {$\sigma$};
			\draw[black] (1.8,1.8) node[right] {$\sigma$};
			\draw[gray] (0,1) circle (0.05);
			\draw[gray] (2,1) circle (0.05);
		\end{tikzpicture}
		\caption{$\rho^2\sigma^2$} \label{fig:M2}
	\end{subfigure}
	\caption{The schematic diagram about two boundary conditions relate to the commutativity of $\rho$ and $\sigma$. Here the red line labels the distribution of state $\rho$ and blue dashed line labels the state $\sigma$. For example, figure \ref{fig:M1} shows schematically that the bulk geometry with the boundary field configuration $\rho\sigma\rho\sigma$. }
	\label{p3}
\end{figure}
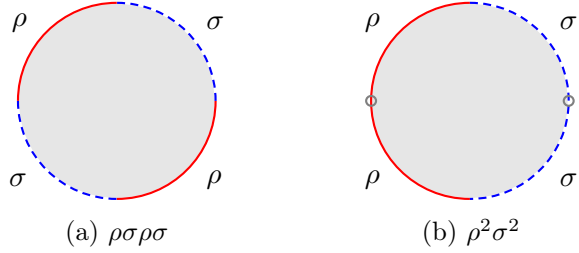

In this section, we will gives a holographic method to check if the density matrices of two holographic states are commutative to each others. In quantum field theory, we can't usually write the density matrix of quantum states precisely. Assume that the quantum state $\rho$ is excited by some operator $\mathcal{O_\rho}(x)$ and other quantum state $\sigma$ is excited by operator $\mathcal{O_\sigma}(y)$. Here ${x,y}$ label the points in the background spacetime. It should be noted that even employing the same operator, the commutator $[\mathcal{O}(x),\mathcal{O}(y)]$ is not necessarily vanish for different spacetime points. However, as mentioned before, there exist a gravity dual for the trace of density matrix's product.
Based on the holographic description, is it possible to determine whether two operators are commutative? In this section, we propose a feasible strategy to answer this question.
Considering two density matrix $\rho$ and $\sigma$ in the same Hilbert space, the commutativity of $\rho$ and $\sigma$ indicates
\begin{equation}
	\rho\sigma=\sigma\rho\,.
\end{equation}
One may think we can use their trace to judge whether they are commutative, but the property of trace $\text{tr}(AB)=\text{tr}(BA)$ rejects this idea. Instead, considering a auxiliary matrix $A$
\begin{equation}
	A=\rho\sigma-\sigma\rho\,,
\end{equation}
if $A=0$ then we can get
\begin{equation}\label{rhosigma}
	\text{tr}(\rho\sigma\rho\sigma)=\text{tr}(\rho^2\sigma^2)\,.
\end{equation}
This result is so straightforward since we can change the position of two density matrix arbitrarily. Conversely given the condition \eqref{rhosigma}, $[\rho,\sigma]=0$ is expected to be proved. Firstly, it is not hard to find that
\begin{equation}
	\text{tr}(\rho\sigma\rho\sigma)=\text{tr}(\rho^2\sigma^2)~\Longrightarrow~\text{tr}(A^2)=0\,.
\end{equation}
Since $A=-A^\dagger$, $A$ can be diagonalized with pure imaginary eigenvalue
\begin{equation}
	A=\sum_{n}ia_n|n\rangle \langle n|,~a_n\in \mathbb{R}
\end{equation}
$\left\{|n\rangle\right\}$ is a complete orthonormal set. $\text{tr}(A^2)=0$ indicates that $a_n=0$ for all $n$, by virtue of anti-hermiticity  we actually obtain that
\begin{equation}
	\text{tr}(\rho\sigma\rho\sigma)=\text{tr}(\rho^2\sigma^2)~\Longleftrightarrow~[\rho,\sigma]=0\,.
\end{equation}
From the perspective of holography, there are two bulk geometries corresponding to boundary condition $\rho\sigma\rho\sigma$ or $\rho^2\sigma^2$, as depicted in Fig.~\ref{p3}. In order to determine whether states $\rho$ and $\sigma$ are commutative, we should calculate these bulk geometries' partition function respectively. Recall that under the saddle point approximation, the partition function of bulk geometry can be written as the exponential function of corresponding action. So we only need to check whether their corresponding geometric action are equal, i.e.
\begin{equation}
	S_{\rho\sigma\rho\sigma}\overset{\text{?}}{=}S_{\rho^2\sigma^2}\,.
\end{equation}
Furthermore, in probe limit we only need to compare the sector of scalar field,
\begin{equation}
	\tilde{I}_{\rho\sigma\rho\sigma}\overset{\text{?}}{=}\tilde{I}_{\rho^2\sigma^2}\,.
\end{equation}
Without losing generality, we continue to use the \added{scalar field} model in previous section. In order to compute the fidelity between $\rho$ and $\sigma$, we already calculate the action of bulk gravity that corresponds to the boundary field $(\rho\sigma)^n$. For a detailed discussion, we will present such result in the previous section again
\begin{equation}\label{rsrs}
	\begin{aligned}
		\tilde{I}_{\rho\sigma\rho\sigma}&=\frac{\beta\Omega_{0,2}}{8\pi G}\left[\frac{\rho_v^2\phi_s^{(0)}}{4\phi_v^{(0)}}+\sum_{j=1}^{\infty}\frac{(1-\cos j\pi)^2}{j^2\pi^2}\frac{\rho_v^2\phi_s^{(j)}}{2\phi_v^{(j)}}\right]\,.
	\end{aligned}
\end{equation}
Here, we need to set $n=2$ to recover $\tilde{I}_{\rho\sigma\rho\sigma}$ from the expression~\eqref{sumI1}. Then we need to calculate the action $\tilde{I}_{\rho^2\sigma^2}$ for boundary field configuration $\rho^2\sigma^2$. Following the same process in previous section, we calculate the action $\tilde{I}_{\rho^2\sigma^2}$ but with boundary condition that is different from~\eqref{boundary} in $n=2$
\begin{equation}
	\tilde{\phi}_v(\tau)=\left\{\begin{aligned}
		\rho_v\quad 0<\tau\leq2\beta \\
		\sigma_v\quad2\beta<\tau\leq4\beta \\
	\end{aligned}
	\right\}\,.
\end{equation}
It is straightforward that period $\tilde{\tau}_{\rho^2\sigma^2}$ is two times than $\tilde{\tau}_{\rho\sigma\rho\sigma}$. And it can be shown that $\tilde{I}_{\rho^2\sigma^2}$ has the same form as $\tilde{I}_{\rho\sigma\rho\sigma}$
\begin{equation}\label{rrss}
	\begin{aligned}
		\tilde{I}_{\rho^2\sigma^2}&=\frac{\beta\Omega_{0,2}}{8\pi G}\left[\frac{\rho_v^2\phi_s^{(0)}}{4\phi_v^{(0)}}+\sum_{j=1}^{\infty}\frac{(1-\cos j\pi)^2}{j^2\pi^2}\frac{\rho_v^2\phi_s^{(j)}}{2\phi_v^{(j)}}\right]\,.
	\end{aligned}
\end{equation}
However, it should be noted that $\phi_s^{(j)}/\phi_v^{(j)}$ has different values in formula~\eqref{rsrs} and \eqref{rrss}. More specifically, we know that $w^{(j)}_{\rho\sigma\rho\sigma}=2\pi j/\tilde{\tau}_{\rho^2\sigma^2}$ and
\begin{equation}
	\tilde{\tau}_{\rho^2\sigma^2}=2\tilde{\tau}_{\rho\sigma\rho\sigma}=4\beta\,.
\end{equation}
Then we can obtain that $w^{(j)}_{\rho\sigma\rho\sigma}=2w^{(j)}_{\rho^2\sigma^2}$. In previous section, we already get the value of $\phi_s^{(j)}/\phi_v^{(j)}$ for $\tilde{I}_{\rho\sigma\rho\sigma}$ and $j$ can only take the odd term, i.e. $j=1,3,5,7,\cdots$. Since $w^{(j)}_{\rho\sigma\rho\sigma}=2w^{(j)}_{\rho^2\sigma^2}$, based on radial equation~\eqref{radial} for $\tilde{I}_{\rho\sigma\rho\sigma}$, we only take $j=\frac{1}{2},\frac{3}{2},\frac{5}{2},\frac{7}{2},\cdots$ to get the exact value of $\tilde{I}_{\rho^2\sigma^2}$. Why can we do this because $\phi_s^{(j)}/\phi_v^{(j)}$ in radial equation~\eqref{radial} changes continuously with the continuous value of parameter $j$. Considering the thermal AdS background, $\tilde{I}_{\rho\sigma\rho\sigma}$ is related to the sum of series~\eqref{defxi1}
\begin{equation}
	\Xi_{\rho\sigma\rho\sigma}:=\sum_{j=1,3,5,\cdots}^{\infty}\frac{\phi_s^{(j)}}{2\phi_v^{(j)}}=-\frac{\sin[\pi(\Delta-\frac{d}{2})](\Delta-\frac{d}{2})\Gamma(\Delta-\frac{d}{2})^2}{2\pi (w_j/2)^{2\Delta-d}}\,.
\end{equation}
According to the above discussion, $\Xi$ for $\tilde{I}_{\rho^2\sigma^2}$ can be easily seen that
\begin{equation}
	\Xi_{\rho^2\sigma^2}=\frac{\Xi_{\rho\sigma\rho\sigma}}{2^{2\Delta-d}}\,.
\end{equation}
So we conclude that $\tilde{I}_{\rho\sigma\rho\sigma}\neq\tilde{I}_{\rho^2\sigma^2}$ and $\rho, \sigma$ are not commutative to each other in our case\footnote{It is noted that $\Xi_{\rho^2\sigma^2}=\Xi_{\rho\sigma\rho\sigma}$ if $\Delta=d/2$. However, as we claimed before, the mass parameter should satisfy $m_\text{BF}^2<m^2<m_\text{BF}^2+1/L^2$ if both $\tilde{\phi}_s(\tau)$ and $\tilde{\phi}_v(\tau)$ can be considered as the source.}.
Another example is considering the Maxwell field $F_{\mu\nu}$ probing the thermal states which are corresponding to AdS-Schwarzschild black hole. In the probe limit, Maxwell field $F_{\mu\nu}$ is decoupled with gravity sector $g_{\mu\nu}$. Therefore, the only different contribution of $\text{tr}(\rho\sigma\rho\sigma)$ and $\text{tr}(\rho^2\sigma^2)$ is the contribution from the Maxwell filed:
\begin{equation}
	I_{\text{Maxwell}}=-\frac{1}{16\pi G}\int_{\mathcal{B}}d^{d+1}x\sqrt{g}F_{\mu\nu}F^{\mu\nu}\,.
\end{equation}
As we have calculated in section~\ref{s3}, the field strength is determined by
\begin{equation}
	F_{\mu\nu}=-\frac{Q(\tau)}{r^{d-1}}(d\tau)_\mu\wedge(dr)_\nu,~Q(\tau)=(d-2)\mu(\tau)r_h^{d-2}\,.
\end{equation}
Here $r_h$ is determined by $\pi L^2/d\beta$. In this case, the chemical potential is time dependent. For $\text{tr}(\rho\sigma\rho\sigma)$, the boundary condition of $\mu(\tau)$ is
\begin{equation}
	\mu(\tau)=\left\{\begin{aligned}
		\mu_\rho\quad\quad 0<\tau\leq\beta \\
		\mu_\sigma~\quad\beta<\tau\leq2\beta \\
		\mu_\rho\quad2\beta<\tau\leq3\beta \\
		\mu_\sigma\quad3\beta<\tau\leq4\beta \\
	\end{aligned}
	\right\}\,.
\end{equation}
Here $\mu_\rho$ and $\mu_\sigma$ are constants. Although $\mu(\tau)$ is time dependent, the field strength has the same nonvanishing component $F_{rt}$. The direct calculation for $I_\text{Maxwell}$ for both $\text{tr}(\rho\sigma\rho\sigma)$ and $\text{tr}(\rho^2\sigma^2)$ reads
\begin{equation}
	\begin{aligned}
		I_{\text{Maxwell}}&=-\frac{1}{16\pi G}\int_{\mathcal{B}}d^{d+1}x\sqrt{g}F_{\mu\nu}F^{\mu\nu}\\
		&=-\frac{2(d-2)\beta\Omega_{k, d-1}(\mu_\rho^2+\mu_\sigma^2)r_h^{d-2}}{8\pi G}\,.
	\end{aligned}
\end{equation}
So we check the consistency of conclusion from section~\ref{s3} in the probe limit, which shows any two charged thermal state in the same Hilbert space are commutative.
As shown in the previous two examples, this gives us a universal\footnote{At least in the probe limit.} holographic approach to check if the density matrices of two holographic states are commutative to each other. As we investigated in this section, the commutativity of $\rho$ and $\sigma$ are holographically determined by the difference between two bulk geometries dual to $\rho\sigma\rho\sigma$ or $\rho^2\sigma^2$.
\section{Summary}
To summary, this paper uses the holographic approach to calculate the fidelity and relative entropy in some concrete examples. By employing the replica tick, we can analytically calculate these quantum distance between thermal states. The key point here is that for two thermal states, the product of their respective density matrices is still a thermal state. In gravity side, this indicates that the gravity dual of the replicated thermal
state is continue to be Schwarzschild-AdS black hole. We also investigate our approach under the presence of matter field.
By the virtue of $U$(1)
invariance for the Maxwell field, these conclusions for the thermal state can be generalized into the charged system within the grand canonical ensemble. The similar step can be also generalized into other cases. Without loss of generality, an Einstein-scalar theory is constructed here to gain the insight into the process of calculating the fidelity. For bulk matter, the replicated boundary field will cause the inhomogeneous boundary condition for
the equation of motion. In general, the physical system we consider here will be time-dependent since the bulk scalar field has inhomogeneous boundary condition. For simplification, we ignore the backreaction from the scalar field to the metric. In other words, the question becomes how to solve the scalar field under the background of fixed Schwarzschild-AdS black hole. Then, we solve the corresponding scalar field's equation of motion in the probe limit. According to the standard holographic renormalization, we derive the analytic expression of the fidelity. The contribution from the scalar operator is a sum of an infinite series. Then we give a numerical result to intuitively feel this contribution here. It is found that on-shell action is divergent under the standard quantization scheme while convergent under the alternative quantization scheme. We observed that this divergency is caused by high frequency modes in Fourier space. While these UV contributions from high frequency modes are suppressed under the alternative quantization scheme. We have concluded that it is a quite common behavior in terms of the bulk matter under inhomogeneous boundary condition. At the end of this paper, a holographic method is discovered to check whether the density matrices of two holographic states are commutative. As examples, we use scalar and Maxwell field to demonstrate this method in the probe limit. The result is expected that two holographic states excited by scalar operator are generally not commutative and two charged thermal states are commutative. However, it should be noted that we do not really solve the Einstein field equation with full backreaction from time-dependent chemical potential. To the best of our knowledge, the exact bulk solution for time-dependent chemical potential is still unclear and worth to explore.
Although this paper studies the fidelity between the states that are excited by a scalar operator, it is also interesting to generalize the procedure into cases of other operators. Further, it is meaningful to generalize our method to study the quantum distance in the subsystem that is a more practical physical system. As an important future problem, more general regularization under the standard quantization scheme is needed to be explore. In the presence of the matter field, the probe limit is very critical for our calculation. In principle, someone can numerically calculate the quantum distance without neglecting the backreaction of the matter field. What is more, the divergency here though is also a UV divergency, it is not caused by AdS boundary and so the usual holographic renormalization cannot not be used to remove such divergency. We leave this problem in the future and call for deep understanding.

\appendix
\section{Euclidean Schwarzschild-AdS$_{d+1}$ black hole}\label{appendix A}
In $(d+1)$-dimensional AdS spacetime, the metric for neutral black is given by
\begin{equation}\label{metric}
	ds^2=-f(r)dt^2+\frac{dr^2}{f(r)}+\frac{r^2}{L^2}d\Sigma^2_{k,d-1}
\end{equation}
with
\begin{equation}
	f(r)=\frac{r^2}{L^2}+k-\frac{f_0}{r^{d-2}}\,.
\end{equation}
Where $L$ denote the AdS radius and $(d-1)$-dimensional metric $d\Sigma^2_{k,d-1}$ is defined as
\begin{equation}
	d \Sigma_{k, d-1}^{2}= \begin{cases}L^2d \Omega_{d-1}^{2}& \text { for } k=+1 \\ d \ell_{d-1}^{2}\equiv\sum_{i=1}^{d-1} d x_{i}^{2} & \text { for } k=0 \\ L^2d \Xi_{d-1}^{2}& \text { for } k=-1\end{cases}\,.
\end{equation}
The parameter $k={-1,0,+1}$ corresponds hyperbolic, planar and spherical horizon geometries respectively. Where $d\Omega_{d-1}^2/d\ell_{d-1}^{2}/d \Xi_{d-1}^{2}$ is the unit metric on $(d-1)$-dimensional spherical/planar/hyperbolic space. In the following discussion, $\Omega_{k, d-1}$ will be denoted as the dimensionless volume of $d\Sigma_{k, d-1}^2/L^2$. Here $f_0$ is interpreted as the mass parameter since that is related to the mass of black hole~\cite{Emparan:1999pm}
\begin{equation}
	M=\frac{(d-1)\Omega_{k, d-1}}{16\pi G_N}f_0\,.
\end{equation}
One can also get the following relation from the blackening factor $f(r_h)=0$
\begin{equation}
	f_0=r_h^{d-2}\left(\frac{r_h^2}{L^2}+k\right)\,.
\end{equation}
The Euclidean Schwarzschild solution is obtained from the ordinary Schwarzschild metric by operating the Wick rotation $t\rightarrow -i\tau$. In order to regulate the conical singularity located at $r_h$, the imaginary time $\tau$ should be identified as
\begin{equation}
	\tau=\tau+\beta
\end{equation}
here the period $\beta=1/T$ under the natural unit $k_B=1$. $T$ is the temperature of the black hole, which is given by
\begin{equation}\label{tem}
	T=\frac{1}{4\pi}\left.\frac{\partial f}{\partial r}\right|_{r=r_h}=\frac{1}{4\pi r_h}\left(d\frac{r_h^2}{L^2}+k(d-2)\right)\,.
\end{equation}

\section{Details about the solution for radial equation}\label{appendix D}
After variable separation, we obtain the following equation, i.e. radial equation\eqref{radial}
\begin{equation}
	R_j''+\left(\frac{f'}{f}-\frac{d-1}{z}\right)R_j'-\left(\frac{m^2}{z^2f}+\frac{w_j^2}{f^2}\right)R_j=0\,, f(z)=1-z^d\,.
\end{equation}
In this paper, we assumes horizon locates at $z=z_h=1$ and $w_j=\frac{dj}{2n}$. At the horizon
we have following asymptotic solution
\begin{equation}
	R_j=a_1(1-z)^{w_j/d}(1+\cdots)+a_2(1-z)^{-w_j/d}(1+\cdots)\,.
\end{equation}
Near the AdS boundary we have following asymptotically behavior:
\begin{equation}
R_j=\phi_s^{(j)}z^{d-\Delta}(1+\cdots)+\phi_v^{(j)}z^{\Delta}(1+\cdots)\,.
\end{equation}
The horizon at the Euclidean black hole will be a smooth tip and scalar field should be
finite here. Thus, we then have following boundary condition
\begin{equation}
	a_2=0\,.
\end{equation}
In subsection~\ref{num}, we need to determine whether the series $\Xi[\rho_v,1/2]$ converges in different quantization schemes. According to comparison test, we obtain that
$$
\frac{\phi^{(j)}_s}{\phi^{(j)}_v}-\left.\frac{\phi_s^{(j)}}{\phi_v^{(j)}}\right|_{\text{thermal AdS}}=\mathcal{O}(j^a)\,, a=-3.81678\approx-4\,.
$$
We note that the numerical result of $a$ is very close to $-4$. In fact, we can analytically confirm this through making a variable transformation for  radial equation,
\begin{equation}\label{ztos}
	s=\int_{0}^{z}\frac{w_j}{f(x)}dx\,,~~R_j=Z_j(s)z(s)\,.
\end{equation}
Then the radial equation is reduced into
\begin{equation}\label{eqz}
	\frac{d^2}{ds^2}Z(s)-\left[V(s)+1\right]Z(s)=0\,,~~V(s)=\frac{f(z)\left[m^2+(d-1)f(z)-zf'(z)\right]}{w_j^2z^2}
\end{equation}
where $z$ is the function of $s$ defined by Eq.~\eqref{ztos} and the boundary condition at horizon becomes
\begin{equation}\label{boundz}
	Z(s)\rightarrow0\,,~~s\rightarrow\infty\,.
\end{equation}
From the definition~\eqref{ztos} we see that $s=w_jz+\mathcal{O}(z^{d+1})$, $f(z)=1-(s/w_j)^d+\mathcal{O}(s/w_j)^{d+1}$ and
\begin{equation}\label{phisv}
	Z[s]=\phi_v^{(j)}w_j^{1-\Delta}s^{\Delta-1}(1+\cdots)+\phi_s^{(j)}w_j^{\Delta+1-d}s^{d-\Delta-1}(1+\cdots)
\end{equation}
near the AdS boundary.
In principle, Eq.~\eqref{eqz} has no analytical solution. However, we can find an analytically approximated solution, which approaches to the exact solution when $w_j$ becomes large enough. We note that
\begin{equation}
	V(s)\rightarrow V^{(0)}(s)=\frac{m^2+d-1}{s^2}
\end{equation}
when $w_j\gg1$. We then use $V^{(0)}(s)$ to replace the potential $V(s)$ and obtain the analytical solution
\begin{equation}\label{an}
	Z_j^{(0)}(s)=b_1\sqrt{s}I_{\Delta-d/2}(s)+b_2\sqrt{s}K_{\Delta-d/2}(s)\,.
\end{equation}
Here $I_v(s)$ and $K_v(s)$ are the first and second kinds of modified Bessel functions. To match the boundary condition~\eqref{boundz}, we have to set $b_1=0$. For small $s$, we have
\begin{equation}
	2K_v(s)=\Gamma(v)\left(\frac{s}{2}\right)^{-v}(1+\cdots)+\Gamma(-v)\left(\frac{s}{2}\right)^{v}(1+\cdots)
\end{equation}
if $v$ is not an integer. We then have
\begin{equation}
	Z^{(0)}_j(s)=\frac{b_2}{2}\left[\Gamma(\Delta-d/2)2^{\Delta-d/2}s^{2-\Delta}(1+\cdots)+\Gamma(d/2-\Delta)2^{d/2-\Delta}s^{\Delta-1}(1+\cdots)\right]
\end{equation}
for small $s$. Using the relationship $s=w_jz+\mathcal{O}(z^{d+1})$ for small $z$ and relationship~\eqref{phisv}, we then obtain
\begin{equation}
	\frac{\phi_s^{(j)}}{\phi_v^{(j)}}=-\frac{\sin[\pi(\Delta-\frac{d}{2})](\Delta-\frac{d}{2})\Gamma(\Delta-\frac{d}{2})^2}{\pi (w_j/2)^{2\Delta-d}}\,.
\end{equation}
For finite but large $c_n$, we can find that $z=s/w_j+\mathcal{O}(1/w_j^{d+1})$ and so
\begin{equation}
	V(s)=V^{(0)}(s)+\frac{(m^2+d)s^{d-2}}{w_j^d}+\cdots\,.
\end{equation}
This shows that
\begin{equation}\label{wj}
	\frac{\phi_s^{(j)}}{\phi_v^{(j)}}=-\frac{\sin[\pi(\Delta-\frac{d}{2})](\Delta-\frac{d}{2})\Gamma(\Delta-\frac{d}{2})^2}{\pi (w_j/2)^{2\Delta-d}}\left[1+\mathcal{O}(w_j^{-d})\right]\,.
\end{equation}
We note that the solution in thermal AdS, i.e. $f(z)=1$ is equal to the analytical solution~\eqref{an} with $V^{(0)}$. In other words, the scalar modes with $w_j\gg1$ can only probe the UV region which is far from the horizon. According to this fact and the above formula~\eqref{wj}, we can get the difference
\begin{equation}
	\frac{\phi^{(j)}_s}{\phi^{(j)}_v}-\left.\frac{\phi_s^{(j)}}{\phi_v^{(j)}}\right|_{\text{thermal AdS}}=\mathcal{O}(1/w_j^{2\Delta})\,.
\end{equation}
So we conclude that $a=-3.81678\approx-4$ is not a coincidence, which is consistent with the previous analysis.
\section*{Acknowledgements}
This work is supported by the Natural Science Foundation of China under Grant No. 12005155.

\bibliographystyle{JHEP}
\bibliography{rose}
\end{document}